\documentclass[preprint,12pt]{elsarticle}
\usepackage{amssymb}
\input epsf

\newtheorem{mytheorem}{Theorem}
\newcommand{\myproof}{\noindent {\bf Proof:\ \ }}
\newcommand{\myqed}{\mbox{\rm Q.E.D.}}

\begin{document}
\bibliographystyle{alpha}
 
\begin{frontmatter}

\title{Candy Crush is NP-hard}
\author{Toby Walsh}
\address{NICTA and University of NSW,
Sydney, Australia}

\begin{abstract}
We prove that playing Candy Crush to
achieve a given score in a fixed number of swaps
is NP-hard. 
\end{abstract}

\begin{keyword}
computational complexity \sep NP-completeness \sep Candy Crush.
\end{keyword}

\end{frontmatter}

\section{Introduction}

Candy Crush Saga is currently the most popular
game on Facebook. It has been installed half a billion times
on Facebook, and on iOS and Android devices. 
%It is 
%estimated that one in
%seven Hong Kong citizens plays the game. 
%Its developer, King.com Ltd is soon to list on the
%NYSE in an initial public offering
%predicted to value the company at 
%over \$5 billion backed by such success. 
Candy Crush is a variant of match-three
games like Bejeweled and Diamond Mine.
These games themselves have also been very popular.
For example, PopGap Games sold over 75 million copies
of Bejeweled. 
But what makes Candy Crush (and its ancestors)
so addictive? In this paper, we suggest one
answer. Namely, part of its addictiveness may 
be that Candy Crush is a computationally hard puzzle to solve.
It thus joins the ranks of other computationally 
intractable puzzles that have fascinated us like 
Minesweeper \cite{minesweeper}, Sudoku \cite{surveypuzzles}, 
and other matching problems like Tetris \cite{tetris}, and KPlumber \cite{kral1}.
It raises a number of open questions.
For example, is the infinite version Turing-complete?
How do we generate Candy Crush problems which are truly
puzzling?

To provide a formal result, we need 
a precise description of the puzzle. 
We focus on the early rounds of Candy Crush Saga
where a player has to achieve a given score with
a fixed number of swaps. A swap interchanges
two neighbouring candies to create a chain
of identical candies. Such chains are deleted
from the board and the candies above drop down
into their place. We also focus on 
simple game play which creates
chains of three identical candies. 
In all the boards we consider, chains
of more than three identical candies cannot
be formed. As in previous work \cite{twipl06}, we consider a generalized
version of Candy Crush in which the board size is not fixed.

\noindent
{\bf Name:} {\sc Candy Crush} problem.

\noindent
{\bf Input:} An $a$ by $b$ board filled with one
of six coloured candies, a number
$k$ of swaps and a score $s$ to be achieved or beaten.
The score equals the number
of chains of 3 identical candies deleted.

\noindent
{\bf Question:} Is there a sequence of $k$ swaps which obtains a 
score of $s$ or more?

In case multiple chains are formed simultaneously, we assume
that chains are deleted from the bottom of the board to the top as 
they appear, and the candies above immediately drop down. 

\section{Result}

%Our main result is that deciding if there is a solution
%to the {\sc Candy Crush} problem is NP-complete. 

\begin{mytheorem}
The {\sc Candy Crush} problem is NP-complete.
\end{mytheorem}
\myproof
A witness is
the sequence of $k$ swaps that results in 
a score of $s$ or more deletions of chains.
The input requires $O(ab)$ bits 
to specify the board, $O(log(k))$ bits
to specify $k$, and $O(log(s))$ bits to
specify $s$. However, specifying the board dominates
the size of the input as both $k$ and $s$ have to
be smaller than $ab$. Each swap %in the witness
can be specified by
giving the coordinates of a square on the
board and a cardinal direction in which to swap it.
The witness is thus $O(k.log(ab))$ bits in size which is
less than the square of the input size. 
Hence, the problem is in NP. 

To show NP-hardness, we reduce an 
instance of 3-SAT in $n$ variables and $m$ clauses
to the {\sc Candy Crush} 
problem. 
The reduction constructs a ``circuit'' using
gadgets and ``wires''.
We set $k=n$. 
We require only 5 of the 6 colours in
the standard {\sc Candy Crush} problem.
In addition, we only ever form chains
of 3 candies of the same colour. We therefore
do not need any special candies like
the ``wrapped'' or ``striped'' candy generated
when longer chains are formed. 
We first show how to construct a neutral 
background that will never result in any chains of 
3 candies of the same colour. In even columns,
we alternate red jelly beans and orange lozenges.
In odd columns, we alternate yellow lemon drops
and green chiclets. We introduce 
various gadgets into this neutral background 
made from purple clusters.
These are inserted in between the red/orange
and yellow/green sequences. The only chains ever formed
will be of purple clusters.
For example, consider the following 
part of the wire
gadget made up of purple clusters:
\begin{center}
\begin{tabular}{ccc}
. & . & . \\
. & {\bf p}  & {\bf p}  \\
{\bf p} & . & .  \\
\end{tabular}
\end{center}
Suppose we insert this into the neutral background:
\begin{center}
\begin{tabular}{ccccc}
r & y & r & y & r \\
o & g & o & g & o \\
r & y & r & y & r \\
o & g & o & g & o \\
r & y & r & y & r 
\end{tabular}
\end{center}
We then obtain the following arrangement of 
candies:
\begin{center}
\begin{tabular}{ccccc}
r & y & r & y & r \\
o & g & o & g & o \\
r & y & {\bf p} & {\bf p} & r \\
o & {\bf p} & r & y & o \\
r & g & o & g & r 
\end{tabular}
\end{center}
This arrangement has the property that, even if we swap candies
around so we get a chain of 3 
purple clusters which then disappears, 
the background colours can never form a chain
of three equal colours. 
For compactness, we describe each
gadget using only a few rows. However, we can
separate apart the different gadgets and even different
parts of a gadget
with neutral rows. 

We next outline the reduction. The board has two parts.
In the left
half of the board, the user
makes choices in setting the
variable gadgets. This
corresponds to setting the respective variables
true or false. In the right half of
the board, we have clause gadgets which 
decide if each clause is satisfied or not. 
We suppose the $i$th variable gadget from the left
represents the truth value of $x_i$. 
The variable gadget contains
two columns of purple clusters.
The user can swap the second
candy down with one horizontally to the left.
This constructs a 
vertical chain of three purple clusters. As a result, 
the middle column of candies moves down
three rows. This corresponds to setting
the variable to false:
\begin{center}
\begin{tabular}{ccc}
. & . & . \\
. & . & {\bf p}  \\
. & . & {\bf p} \\
. & {\bf p} & .  \\
. & {\bf p} & .  \\
. & . & . \\
\end{tabular}
left swap
$\Rightarrow$  
\begin{tabular}{ccc}
. & . & . \\
. & . & {\bf p}  \\
. & {\bf p} & .  \\
. & {\bf p} & .  \\
. & {\bf p} & .  \\
. & . & . \\
\end{tabular}
$\Rightarrow$
\begin{tabular}{ccc}
. &  & . \\
. &  & {\bf p}  \\
. &  & .  \\
. & .  & . \\
. & .  & . \\
. & . & . \\
\end{tabular}
\end{center}
Alternatively, the user can swap the third purple cluster
down with one horizontally to the right. This 
also gives a 
vertical chain of three purple clusters. As a result,
the rightmost column of candies moves down
three rows. This corresponds to setting
the variable to true. 
\begin{center}
\begin{tabular}{ccc}
. & . & . \\
. & . & {\bf p}  \\
. & . & {\bf p} \\
. & {\bf p} & .  \\
. & {\bf p} & .   \\
. & . & . \\
\end{tabular}
right swap
$\Rightarrow$  
\begin{tabular}{ccc}
. & . & . \\
. & . & {\bf p}  \\
. & . & {\bf p}  \\
. & . & {\bf p}  \\
. & {\bf p} & .   \\
. & . & . \\
\end{tabular}
$\Rightarrow$
\begin{tabular}{ccc}
. & . &  \\
. & . &   \\
. & . &   \\
. & . & . \\
. & {\bf p} & .  \\
. & . & . \\
\end{tabular}
\end{center}
In both cases, exactly one vertical chain
of 3 purple clusters is created. All our gadgets
except for the clause gadget
are constructed so that the same number
of chains of 3 purple clusters are created whatever
setting of truth values are chosen by the user. 
In this way, the biggest change that can be made
to the final score will be by satisfying clauses.
The maximum final score will require us to 
satisfy all the clause gadgets. 
In some cases, we pair gadgets together
so that, irrespective of their inputs,
they construct together the same number of 
chains of 3 purple clusters. 
All gadgets, except the clause
gadgets, are 3 columns wide.

Next, we describe a ``wire''. This will transmit information
across the board. Initially the
input and output contain candies
from the neutral background. If 
a purple cluster 
is placed at the input, then a purple cluster appears shortly
after at the output. 
\begin{center}
\begin{tabular}{cccc}
 & . & . & {\bf p} \\
 & .  & . & $out$ \\
 & {\bf p}  &  {\bf p} & .    \\
 & .  &  . & {\bf p} \\
 &.  &  .   & . \\
&.  &  .   & . \\
$in$ & {\bf p}  & {\bf p} & . 
\end{tabular}
\end{center}
Suppose a purple cluster is introduced at
the input. Then the board goes through the
following changes:
\begin{center}
\begin{tabular}{cccc}
& .        & .& {\bf p} \\
& .        & .& . \\
& {\bf p}  & {\bf p} & . \\
& .  &  . & {\bf p} \\
&.  &  . & .       \\
. &.  &  . & .       \\
{\bf p} & {\bf p}  & {\bf p} & .
\end{tabular}
$\Rightarrow$
\begin{tabular}{cccc}
& .        & . & {\bf p} \\
& .        & . & . \\
& . & . & . \\
& {\bf p}  & {\bf p} & {\bf p} \\
& .  &  . & . \\
. &.  &  . & .       \\
. & .  &  . & .       \\
\end{tabular}
$\Rightarrow$
\begin{tabular}{cccc}
& . & . & .  \\
& . & . & {\bf p} \\
& . & . &  . \\
& . & . &  . \\
& .  &  . & . \\
. & .  &  . & . \\
.  & .  &  . & .       \\
\end{tabular}
\end{center}
We can glue wires together to communicate
information across longer distances. 
Note that we can add any number of neutral rows into 
the middle of such wire gadgets. 
Note also that when a wire communicates
information, two purple clusters
are deleted in each column. 
Wires always come in pairs: one 
representing a literal and the other its
negation. 
By construction, each pair of wires
then deletes two rows from the table.
The gadgets above therefore all 
come down in unison by two rows. 
Hence they remain connected in the
same way as in the initial board.

We also need a connector gadget that connects a wire
to the columns above the variable gadget. 
For example, suppose we want to connect
a wire to the columns above
the $i$th variable to represent
the literal $x_i$. 
The following connector gadget 
creates an output bit that can
act as the input bit to a wire gadget. 
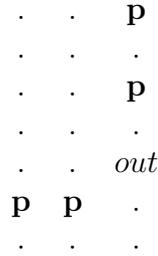
\begin{center}
\begin{tabular}{ccc}
. & . & {\bf p} \\
. & . & . \\
. & . & {\bf p} \\
. & . & . \\
.  & . & $out$ \\
{\bf p} & {\bf p} & . \\
.  &  .   & . 
\end{tabular}
\end{center}
This sits above a variable gadget. 
A dual
construction is used to connect
a wire to represent the
literal $\neg x_i$. 
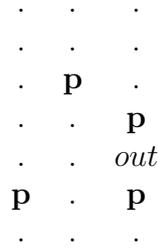
\begin{center}
\begin{tabular}{ccc}
. & . & . \\
. & . & . \\
. & {\bf p} & . \\
. & . & {\bf p} \\
.  & . & $out$ \\
{\bf p} & . & {\bf p} \\
.  &  .   & . 
\end{tabular}
\end{center}
As such connectors come in pairs,
we have different numbers of purple clusters
deleted from the different columns. 
Any connector gadgets above therefore
need additional neutral
candies to compensate. 

We also have wires that cross
columns containing variable gadgets. 
This requires modifying the wire
gadget to deal with the differing shifts of 
the three columns. Suppose we pass
a wire above the variable gadget but beneath
any of connectors. 
Our modified wire gadget is as follows:
\begin{center}
\begin{tabular}{cccc}
 & . & . & {\bf p} \\
 &.  &  .   & . \\
 &.  &  .   & . \\
 & . & . & {\bf p} \\
 & .  & . & ${out}$ \\
 & .  &  {\bf p} & .    \\
 & . & .   &  {\bf p}    \\
 &.  &  .   & . \\
 & {\bf p}  &  {\bf p} & .    \\
 & .  &  . & {\bf p} \\
 & .  &  {\bf p} & .    \\
 &.  &  .   & . \\
&.  &  .   & . \\
$in$ & {\bf p}  & {\bf p} & . 
\end{tabular}
\end{center}
The modification has added additional rows to
the wire gadget to ensure that, whatever
truth value is selected, when the apropriate 
three clusters have been deleted in the variable
gadget, we have 
purple clusters in all the positions
of the original wire gadget. 
Similar modifications of
the wire gadgets are required for 
wires that pass above connectors above
a variable gadget. As before, we modify the
wire gadget to include
additional neutral candies to compensate
for the different number of deleted candies. 

In the right hand part of the board, 
we have a clause section. Each clause is represented
by $m$ clause gadgets. 
This repitition of clause gadgets is 
required to ensure the user gains the maximum
score by setting variable gadgets and not
indirectly by setting individual wires leading
to a single one of these clause gadgets.
Each clause gadget occupies a block of
rows not occupied by any other clause gadget. The clause
gadgets
are arranged diagonally from
top left  to bottom right. 
In this way, no wires need pass over the
top of any clause gadget. We therefore do not
need to worry about the number of 
purple clusters deleted in each column
of the clause gadget. 

Suppose we have the clause $x_3 \vee \neg x_5 \vee x_{19}$. 
From top to bottom, we have six wires
 coming from the variable
section of the board
representing the signal $\neg x_3$, $x_3$, $x_5$, $\neg x_5$,
$\neg x_{19}$, $x_{19}$. We denote these wires
by $\overline{in_3}$, $in_3$, 
 $\overline{in_2}$, $in_2$, 
 $\overline{in_1}$, $in_1$ respectively.
We connect these wires to a clause 
gadget. The bottom part of the clause gadget is
described in Figure 1.

\begin{figure}[p]
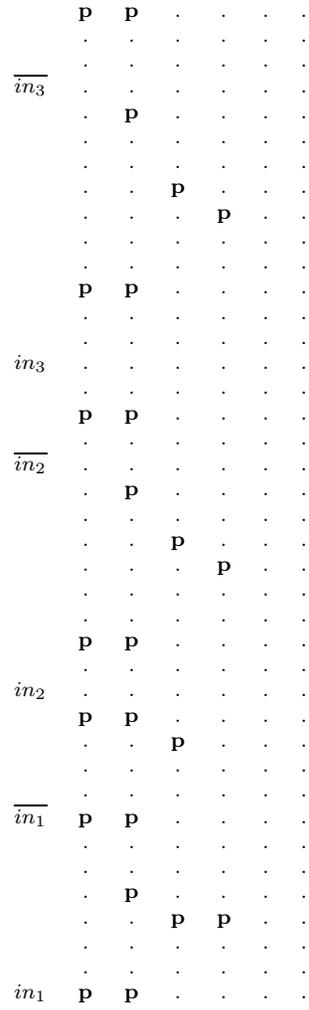

{\scriptsize
\begin{center}
\begin{tabular}{lcccccc}
      & {\bf p} & {\bf p} & . & . & . & .\\
      & . & . & . & . & . & .\\
      & . & . & . & . & . & .\\
$\overline{in_3}$  & . & . & . & . & .& . \\
      & . & {\bf p} & . & . & . & .\\
      & . & . & . & . & .& . \\
      & . & . & . & . & .& . \\
      & . & . & {\bf p} & . & .& . \\
      & . & . & . & {\bf p} & . & .\\
      & . & . & . & . & . & .\\
      & . & . & . & . & . & .\\
      & {\bf p} & {\bf p} & . & . & .& . \\
      & . & . & . & . & . & .\\
      & . & . & . & . & . & .\\
$in_3$ & . & . & . & . & . & .\\
      & . & . & . & . & . & .\\
      & {\bf p} & {\bf p} & . & . & . & .\\
      & . & . & . & . & . & .\\
$\overline{in_2}$  & . & . & . & . & .& . \\
      & . & {\bf p} & . & . & . & .\\
      & . & . & . & . & .& . \\
      & . & . & {\bf p} & . & .& . \\
      & . & . & . & {\bf p} & . & .\\
      & . & . & . & . & . & .\\
      & . & . & . & . & . & .\\
      & {\bf p} & {\bf p} & . & . & .& . \\
      & . & . & . & . & . & .\\
$in_2$ & . & . & . & . & . & .\\
      & {\bf p} & {\bf p} & . & . & . & .\\
      & . & . & {\bf p} & . & .& . \\
      & . & . & . & . & . & .\\
      & . & . & . & . & . & .\\
$\overline{in_1}$ & {\bf p} & {\bf p} & . & . & . & .\\
      & . & . & . & . & . & .\\
      & . & . & . & . & .& . \\
      & . & {\bf p} & . & . & . & .\\
      & . & . & {\bf p} & {\bf p} & . & .\\
      & . & . & . & . & . & .\\
      & . & . & . & . & . & .\\
$in_1$ & {\bf p} & {\bf p} & . & . & .& .
\end{tabular}
\end{center}
}
\caption{Bottom half of the clause gadget.}
\end{figure}
When one or more of the input wires ($in_1$,
$in_2$, or $in_3$) is set, 
two chains of three purple clusters
are created within the clause gadget. 
This occurs when the wires represent
a truth assignment that satisfies the clause. 
As a result, the fourth column from the
left in the clause gadget 
drops down one row if and only if
the clause is satisfied by the truth assignment
set on the input wires. 
In the top half of the clause gadget, when the fourth
column drops, the score is increased greatly. The increase in the 
score is larger than any other score that
might be achieved. In particular, 
satisfying the clause creates $ma$ chains of
purple clusters in the top of the clause
gadget. 
As each clause gadget is repeated
$m$ times, satisfying all the clause gadgets
by assigning variables out scores
anything else that the user can do. For example,
the user can set wires directly, even set a wire
and its negation. However, this will score
less than just setting the variable
gadgets provided this satisfies all the clause gadgets. 
The 
top half of the clause gadget has $ma$ copies
of the following purple clusters stacked above each other:
\begin{center}
\begin{tabular}{lcccccc}
      & . & . & . & . & . & .\\
      & . & . & . & {\bf p} & . & .\\
      & . & . & . & . & {\bf p} & {\bf p} \\
      & . & . & . & . & . & .
\end{tabular}
\end{center}
Directly setting a wire will 
create less than $a$ chains of purple 
clusters as this is the length of the longest possible
wire. This is less productive than satisfying
a clause. In fact, the only way we can get the maximum achievable score
is to set only variable gadgets with an assignment that satisfies
all the clause gadgets. Hence, deciding if we can achieve
the (given) maximum achievable score determines the satisfiability
of the original 3-SAT problem.
Finally, we note that the reduction is polynomial. The board is 
$O(n+m^2)$ wide and $O(m^3(n+m^2))$ high.
\myqed

Note that the proof only creates chains of 
3 identical candies. 
Other closely matching problems in the same family
like (generalized versions of) Bejeweled and Diamond Mine are 
therefore also NP-complete. 
The proof required a board that was 
unbounded in width and height. 
When we bound the width (or height) of the
board, is the problem fixed parameter tractable. 
Another interesting open question is if we can
approximate the problem easily. For instance, can
we get within a constant factor of the optimal
score? Or can we solve
a problem within few additional swaps? 

Note also that the whole board was
visible and playable. In many levels of Candy Crush, only 
the bottom part of the board is visible to 
the player. %We therefore only have
%partial information about the board.
The problem of playing Candy Crush
remains NP-hard with partial information (since
a special case of partial information is when 
there are no scores from candies that
fall down from the hidden part of the
board into the playable and visible section). 
%Indeed, it is not hard to see that we 
%will move up the polynomial hierarchy when we consider, 
%for example, swaps on the visible part of the
%board that lead to a solution
%irrespective of how the hidden candies
%are ordered, or that are most likely
%to lead to a solution supposing some
%distribution of the hidden candies. 
%
Finally, puzzles like Candy Crush may not be very puzzling
if they have many solutions. 
%We can also
%show that 
Deciding if a {\sc Candy Crush} problem
has an unique solution is co-NP-hard.
%We formally define the following problem.

%\noindent
%{\bf Name:} Unique {\sc Candy Crush} problem.

%\noindent
%{\bf Input:} An $a$ by $b$ board of candies, a number
%$k$ of swaps and a score $s$. 

%\noindent
%{\bf Question:} Is there an unique sequence of $k$ swaps which
%obtains a score of $s$ or more?

\begin{mytheorem}
The Unique {\sc Candy Crush} problem is co-NP-hard. 
\end{mytheorem}
\myproof
We reduce from the complement of the unique SAT problem.
The unique SAT problem is itself co-NP-hard. 
We use a similar reduction as in the last proof.
We modify the variable gadgets so that each %variable gadget 
only works when an input wire is set and 
so that each variable gadget
sets an output wire when the user has selected
a variable asssignment. We then
wire the variable gadgets up from left
to right. In this way, we ensure that the variable gadgets
must be set in sequence from left to right.
This ensures that we cannot permute the order
in which we set the variable gadgets.
The SAT instance has an unique solution if and only if
there is an unique sequence of variable
assignments achieving the (given) maximum score.
\myqed

\section{Conclusions}

We have shown that the generalized version of Candy Crush
is NP-hard to play. 
This result suggests a number of interesting future
directions.
For example, NP-hardness is only a worst-case concept.
How do we generate Candy Crush 
problems that are hard in practice? 
Can we identify a ``phase transition'' in hardness?
Phase transitions have been seen in many
other NP-hard problems and are now frequently used
to benchmark new algorithms
\cite{cheeseman-hard,easy-hard,rnp,GentIP:tsppt,wjair11}.
How does the structure of Candy Crush problems
influence their hardness \cite{gwhcp06}?
Finally, it would be interesting to see if 
we can profit from the time humans spend solving
Candy Crush problems. Many 
millions of hours have been spent solving Candy Crush. Perhaps
we can put this to even better use by 
hiding some practical NP-hard problems
within these puzzles?

\bibliography{/Users/twalsh/Documents/biblio/a-z,/Users/twalsh/Documents/biblio/a-z2,/Users/twalsh/Documents/biblio/pub,/Users/twalsh/Documents/biblio/pub2}
%\bibliography{/Users/twalsh/Documents/biblio/a-z2,/Users/twalsh/Documents/biblio/pub2}
%

\end{document}